\documentclass{elsart}
\usepackage{graphicx} 
 
\begin{document} 
 
\title{Amplified imitation in percolation model of stock market} 
 
{\Large D.~Makowiec\footnote{fizdm@univ.gda.pl}, P. Gnaci\'nski and W.~Miklaszewski} 
\address{Institute of Theoretical Physics and Astrophysics,  
Gda\'nsk University \\ ul.Wita Stwosza 57, 80-952~Gda\'nsk, Poland} 
 
\maketitle

{\small The herd behavior of the Cont Bouchaud model is amplified by allowing
clusters to copy decisions of some other cluster in the next time step.
The results of the model are compared to data from Warsaw Stock Exchange.
It follows that the mechanism  of the amplified imitation could 
be responsible for the sell decision on a poorly developed, emergent market. 
} 
 
 
 
\baselineskip = 18pt 
 
\section{Introduction} 
Over the last decade, artificial economic world living in computer made
markets of  financial assets  has become an interest
for physicists experienced in complex systems research. See
\cite{Levy,Samanidu} for review. 
The reasons for the Cont-Bouchaud proposition is to explain effects of herd
behavior which is observable  among real traders.  Differently
from models of  markets with  agents   rational, or
heterogeneous, or adaptive,  the Cont-Bouchaud model (CB-model, in short) considers noisy
traders only, in the sense that traders do not have any rational strategy. Instead the strong communication and information links between traders is proposed. In effect agents form groups- called clusters, within which behavior of agants is common.

Complex system approach provides percolation as a powerful tool in study
of connected objects of different type and origin,  
\cite{AharonyStauffer,StanleyAndrade}.  The Cont-Bouchaud  proposition is to adapt the percolation phenomena to design a  stock market organization \cite{ContBouchaud}. The original model uses infinite-range bond percolation. Similar interactions are proposed by  considering nearest neighbor and site percolation instead \cite{StaufferSornette,StaufferPenna,CastiglioneStauffer} At a very few assumptions, the model reproduces, so-called stylized, facts known about fluctuations of price of real market assets which are widely accepted, \cite{Cont}. In particular, the price fluctuations  measured by so-called returns, i.e., change of the market price, (i) are not correlated, (ii) average of them is zero and (iii) tails of a distribution of returns decay accordingly to the power-law. These features are achieved only due to exploration of critical properties attributed to the percolation phenomena.
 
In the CB-model  there is no other interaction between investors
than instantaneous imitation through a cluster structure (these structure can change in time \cite{Eguiluz}). Our proposition  is to consider imitation of one cluster  by other clusters  but  taken in the $next$ time step. We say the cluster imitates some other cluster with a time step delay. In this sense the herd behavior of the CB-model  is amplified by the delay imitation. 

The  additional interaction introduced  seems to
be particularly suitable when poorly developed stock markets are considered.
In the result of multi century experience the large markets, like London City
or New York Stock Exchange, have developed the solid structures and their
regulations are firmly executed. 
One can say that the fundamental assumption of Efficient Market Hypothesis
\cite{Fama} that the whole information is available for all investors at the same instant, is close to be satisfied. The race and competition
between individual investors effects in that the same information is differently
understood and in consequence differently transferred into market decisions.
There is no place for  imitation between investors for at least two
reasons. First, the next piece of information arrives at the market and
investors are forced to concentrate on this new subject. Second, investors
are known to be extremely strong individuals in the sense it is said, they are  overconfident about their personal understanding of the market. Such attitude leads to the strong disagreement between investors rather than cooperation\cite{overconfidence}. However when a market is far from equilibrium, because, for example, a market is strongly influenced by political decisions or liquidity of a stock is limited then the only strategy to win is to search for `well informed' investors to join their team or if it is impossible to follow  their decisions. 
 
In the subsequent sections we provide:  setting of the model (Section
2), analysis of  results obtained in simulations of the model (Section 3),
examples of stocks data  from  the Warsaw Stock Exchange market for which the
hypothesis of amplified  imitation as underlying mechanism of driving
the price can be proposed (Section 4). In the closing section further possible 
modifications to the model  are proposed.
 
\section{Details of the model} 

In  the Stauffer {\it et al.} representation  \cite{StaufferSornette,StaufferPenna,CastiglioneStauffer} of the CB-model one deals with a number of investors who are distributed on a square lattice of a linear size $L$. Each of the investors is assigned to a site at random with the probability $p$. A site can be occupied by one investor  or left empty. The geometric relations arising from nearest neighboring occupied sites, called clusters, are interpreted as coalitions of investors-  groups of traders that act together on a stock market. The coalitions last for a limited period of time  after which a new arrangement of stock market participants is introduced.  The clusters take decisions randomly and independently of each other to $buy$ or $sell$, with the probability $a$, or just $wait$ with a probability $1-2a$. 
The resulting stock price is proportional to the sum of  demand and
sell orders from all clusters which have been active during a time step.
The parameter $a$ is called $activity$ and  can be seen as a control of 
the length of a time step \cite{StaufferPenna}. If activity is high, which
means that $a$ is close to $0.5$, then all clusters are active during a session.
Such activity  characterizes a long time step. The independent and stochastic
decisions  taken during long sessions  lead to the histogram of
price changes of a Gaussian type. If the activity is low, $a < {1\over
L^2} $, what means that  a time step is short, then on average one cluster
acts during a given market session. In effect the histogram of  price
changes reflects properties of the cluster size distribution. If $p$ is close
to the percolation threshold then the power-law decay of tails of the distribution is observed with the exponent $\tau =2$ in case of a square lattice \cite{AharonyStauffer}.  
 
Let investors occupy randomly a fraction $p_c=0.592746$ of the sites of $L\times
L$ square lattice. This particular value of the parameter $p_c$ provides
critical properties for a cluster structure on a square lattice: a cluster that spans the lattice occurs with the probability $1\over 2$ . In the following we use the Hoshen-Kopelman algorithm \cite{HoshenKopelman} to identify clusters.

The clusters of size greater than 2 are split into two classes:\\ 
--- class A contains clusters of well informed agents. These clusters take
decisions independently and at random:  they  buy or sell at the
probability $a$ and at probability $1-2a$ they do nothing.\\ 
--- class B  contains clusters which imitate other clusters. Each cluster
is linked to some other cluster. Imitation means that the next time step
decision of a B-class cluster is exactly the same as the present decision
of the cluster to which the B-class cluster is linked. \\ 
Let us assume that all clusters of size smaller than 3 run the ordinary CB-model
dynamics. 
 
\begin{figure} 
\begin{center} 
       \includegraphics[width=0.4\textwidth, trim= 0 0 0 0]{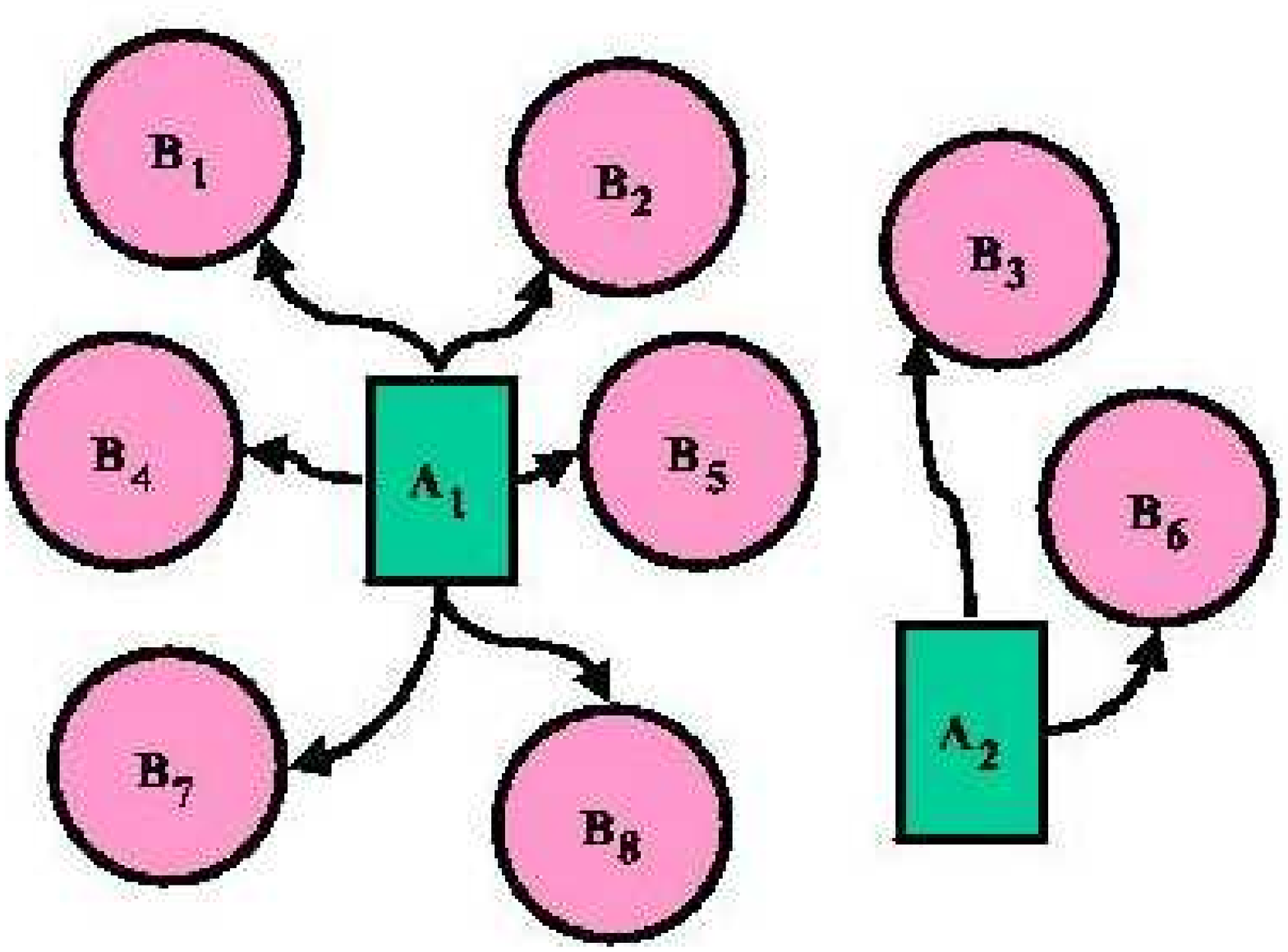}\hspace{1.0cm} 
       \includegraphics[width=0.4\textwidth, trim= 0 0 0 0]{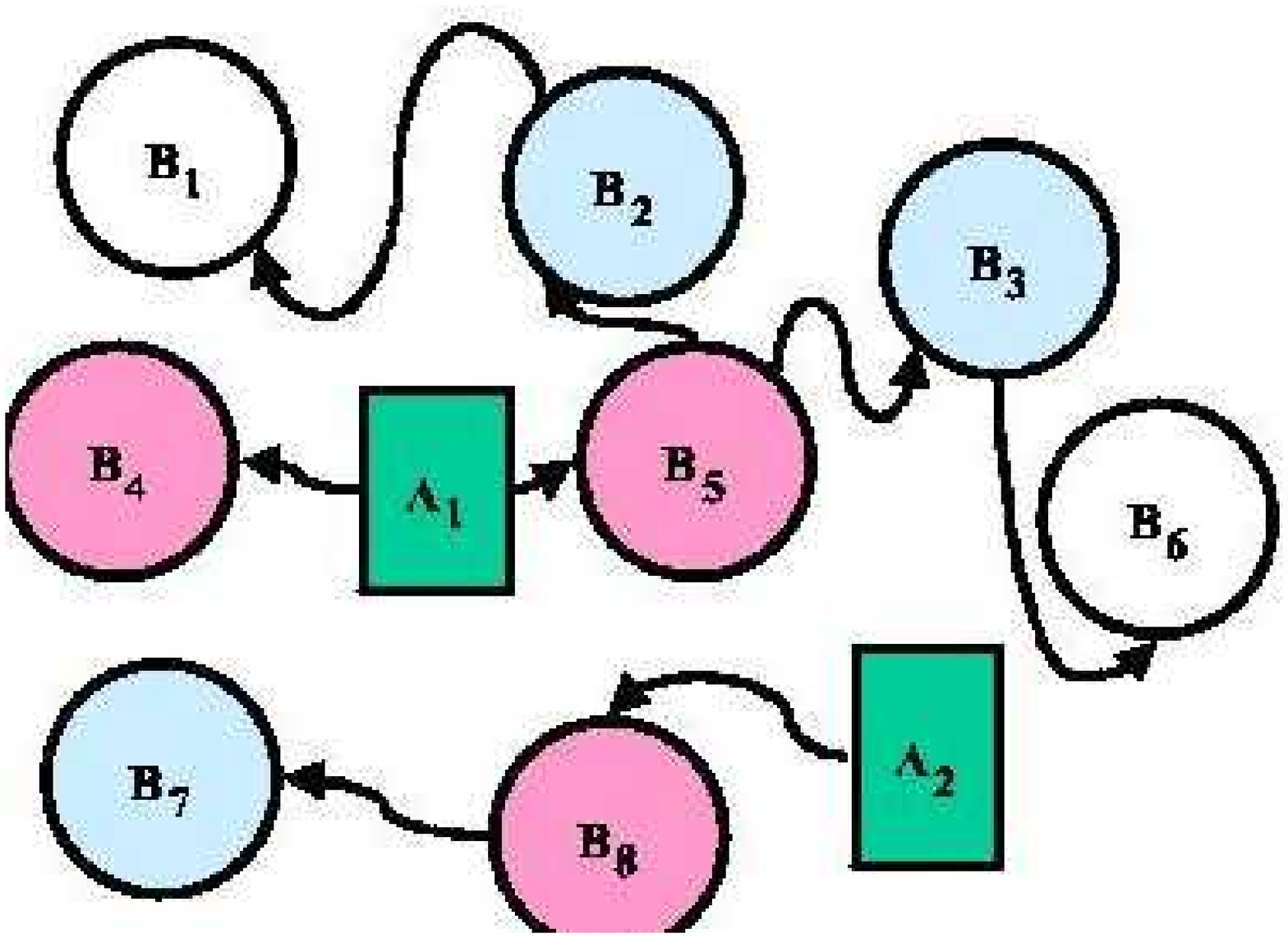} 
\end{center} 
    \caption{The network of imitation between clusters-- an
illustration. Green (dark in black and white view) rectangles represent clusters
of the class A and circles represent clusters of the class AB- less red color
(less gray) means longer delay time in imitation; left panel: trees of the
model A; right panel: trees of the model AB.} 
    \label{fig:schemat} 
\end{figure}

The following two types of imitation will be considered: {\bf model A} -
each cluster from  the class B at random chooses a cluster from the
class A to imitate,  and {\bf model AB}  - each cluster from the
class B is linked to  a randomly choosen cluster from any class. So that the pool of
clusters is arranged into trees, see Fig. \ref{fig:schemat}.
The root of each tree is a leader- an independently acting participant of a market. The
rest of clusters grupped in a tree is either of one step depth tree --- case of the model A, or many steps depth tree --- case of the model AB. Clusters from a tree take the same market decision as their leader but with a delay which  equals to the path length to the leader of the tree.  
 
Let us introduce the parameter $b\in (0,1]$ the probability that a cluster
of size greater than 2 belongs to the class A. At a given $b$ the average
number of clusters in one tree is  
$$ 1+ {(1-b)N\over bN}= {1\over b} $$ 
where $N$ is the average number of clusters of size greater than 2. Hence
if $b=1$ then both models collapse to the original  CB-model.

\section{Results}

Our simulations are performed on a square lattice of the linear size  $L=200$.
On this lattice at $p_c$ the average number of clusters of size greater than
2 is $391\pm 25$ (total number of all clusters of 1's is $1205 \pm 50$). Setting 
$b=0.001$ indicates that at most one tree is present on the lattice while $b=0.005$
($b=0.01 $)  allows  two (four, respectively) trees to exist on the
lattice. If $b=0.1$ then at average  about $40$ trees take part in the market game.
Therefore, very long runs must be considered to obtain satisfactory statistics.
Our simulations are performed  for {\it (number of lattice rearrangements)}
$*$ {\it (length of each of arrangement) }= $10^6 * 200$ time steps and experiments
were repeated several times.

\begin{figure} 
\begin{center} 
        \includegraphics[width=0.7\textwidth, trim= 0 0 0 0]{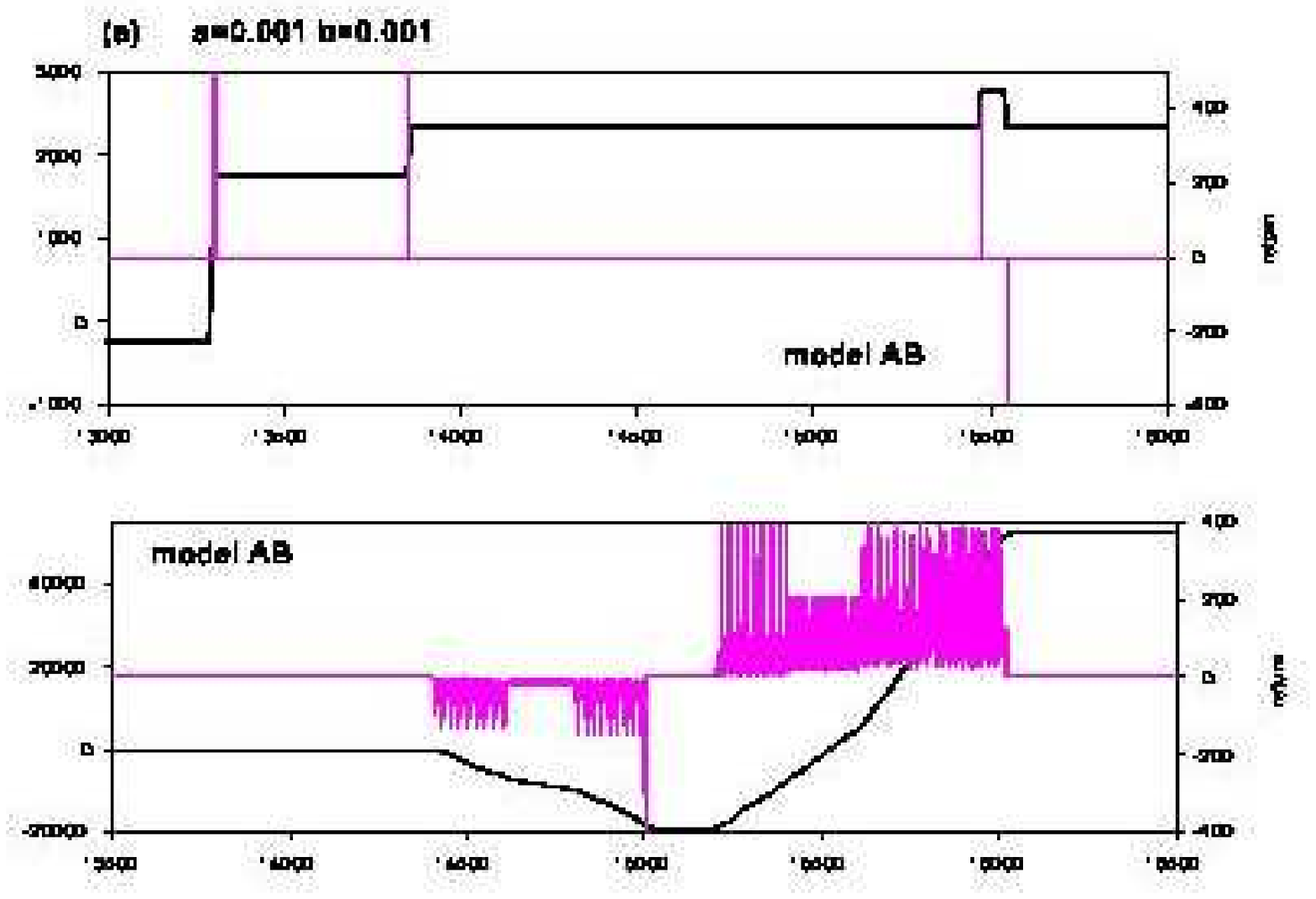} 
        \includegraphics[width=0.7\textwidth, trim= 0 0 0 0]{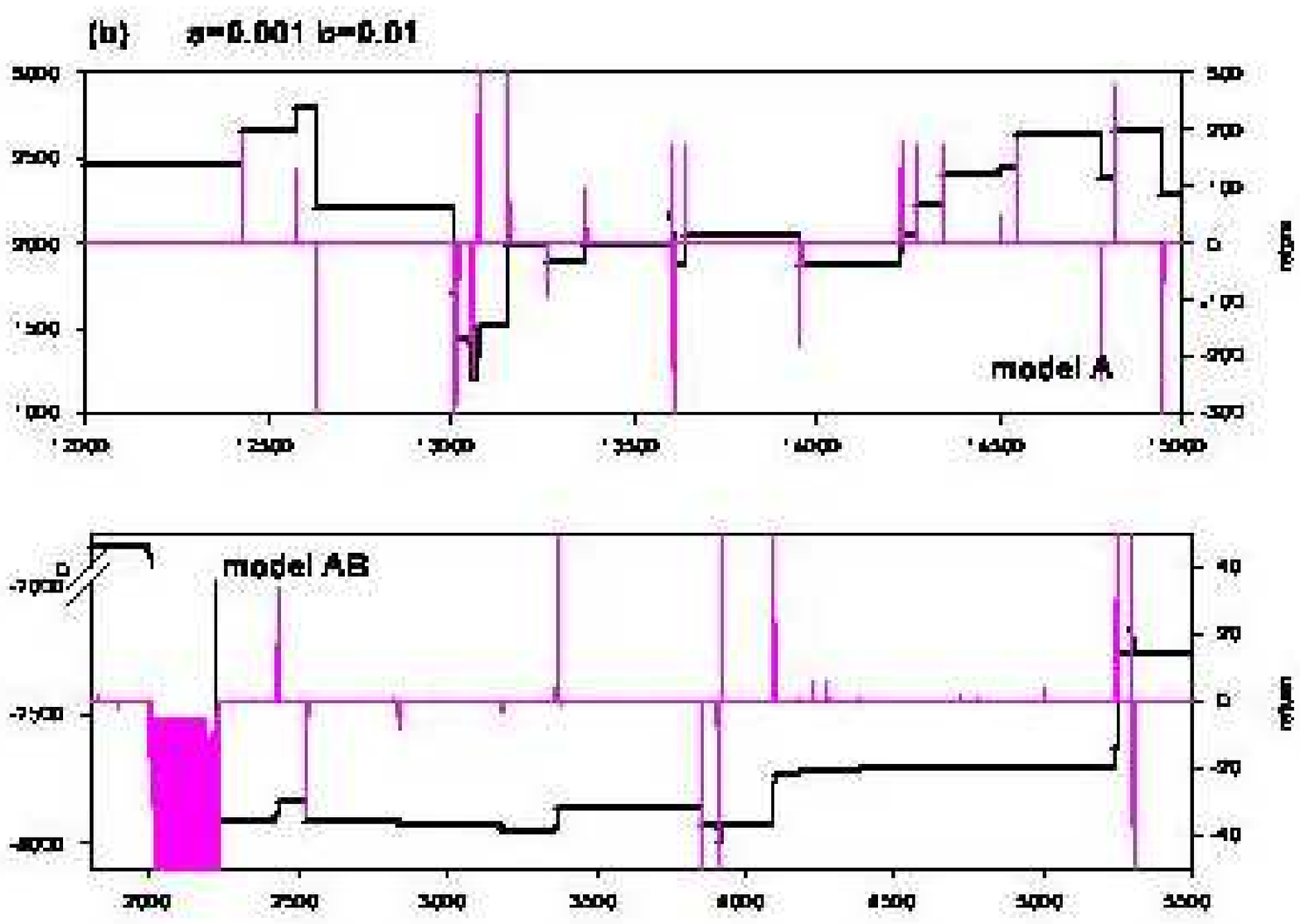} 
        \includegraphics[width=0.7\textwidth, trim= 0 0 0 0]{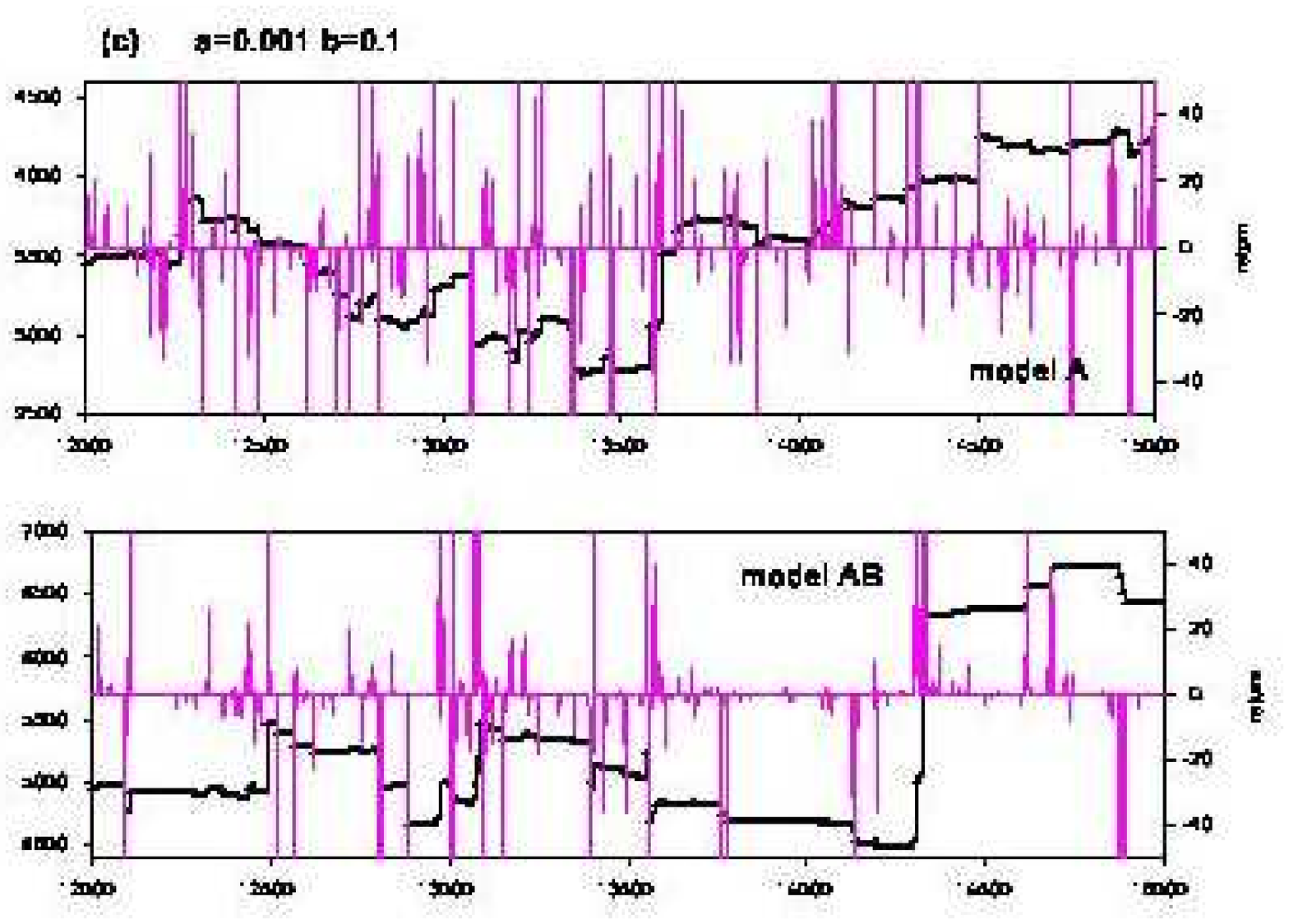} 
\end{center} 
    \caption{Examples of series of  price and returns
arising from both models if (a) $b=0.001$, (b) $b=0.01$, (c) $b=0.1$.} 
\label{fig:series} 
\end{figure} 
 
Let us start with presenting  examples of time series emerging from the two microdynamics
studied, Fig.~\ref {fig:series}. The activity $a=0.001$ in all experiments.
This value of activity has been found as satisfactory low to be sensitive to
observe critical cluster distribution, i.e., the power-law decay of distribution tails. \cite{StaufferPenna,CastiglioneStauffer}.
At average the activity of the whole market is constants, but existing tree 
structures lead to strongly synchronized actions. In effect we observe huge
impacts on a price change after each decision done by a member of the class
A if there are only few trees on a lattice, see Fig.~\ref{fig:series}(a) and
Fig.~\ref{fig:series}(b). In case of the model A  isolated high changes
of a price are presents. In case of the model AB an avalanche of price changes
developes. If the number of leaders grows,  see Fig.~\ref{fig:series}(c),
trees of dependencies are smaller and  values of returns go down. However
we still observe bursts of activity. 
The volatility, by means the standard deviation,  of the recorded data
for different values of $b$ is as following: 
\begin{equation} 
\begin{array}{lrr} 
     b & \qquad model\ A & \qquad model \ AB \\ 
    0.001 &\qquad\qquad 21.2 &\qquad \qquad 31.2 \\ 
    0.005 &\qquad\qquad 38.5 &\qquad \qquad 32.6 \\ 
    0.01 &\qquad\qquad 39.5 &\qquad \qquad  30.7\\ 
    0.1 &\qquad\qquad 21.8 &\qquad \qquad 20.4 \\ 
    0.5 &\qquad\qquad 19.5 &\qquad \qquad 19.4 \\ 
    1.0 &\qquad\qquad 19.3 &\qquad \qquad 19.3 \\ 
\end{array} 
\end{equation} 
So that one can say that for $b > 0.1$ the volatility in  both systems is close to the volatility of the pure CB-model. 
 
\begin{figure} 
    \begin{center} 
        \includegraphics[width=0.6\textwidth, trim= 0 0 0 0]{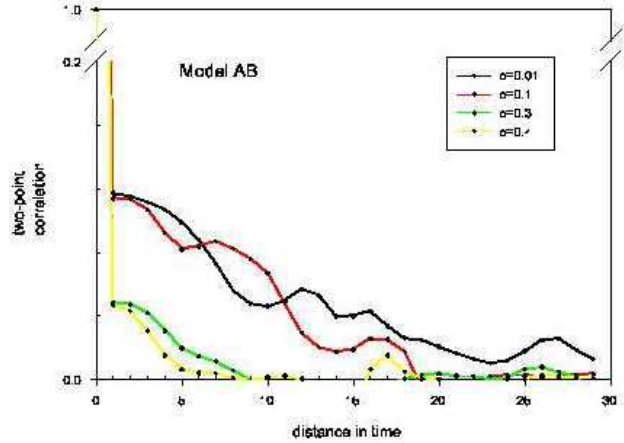} 
    \end{center} 
    \caption{Two-point autocorrelation function  for
the model AB for different values of b} 
    \label{fig:correl} 
\end{figure}

It occurs that the two-point autocorrelation function decays sharply in 1
step independently of $b$ as far as the model A is considered. However in
case of the model AB the presence of delay trees transmits into weak correlation.
We observe  nonzero correlation but $ < 0.2$  for few time steps
if $b < 0.4$, see Fig.~\ref{fig:correl}. 
 
\begin{figure} 
    \begin{center} 
        \includegraphics[width=0.48\textwidth, trim= 0 0 0 0]{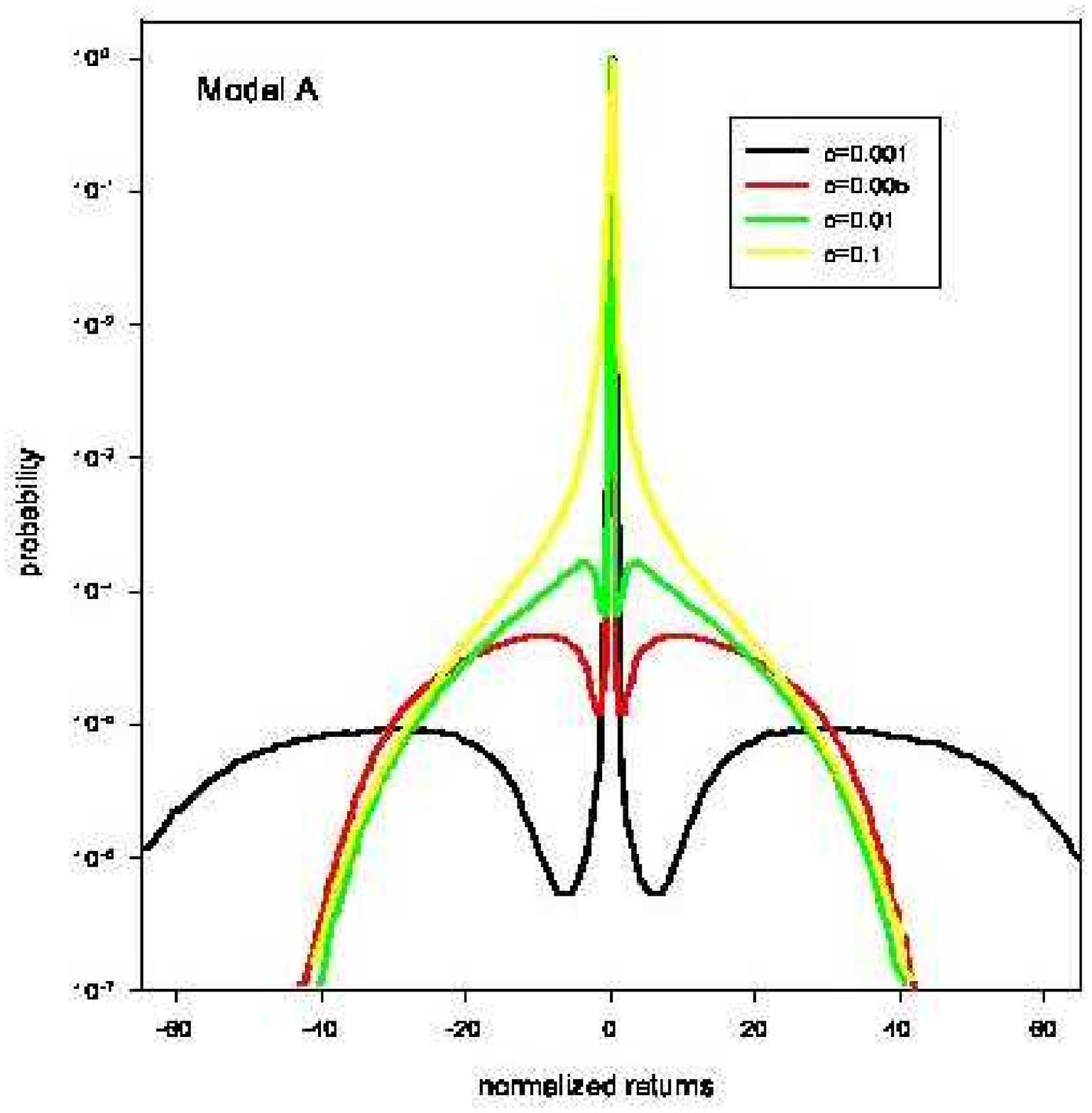} 
        \includegraphics[width=0.48\textwidth, trim= 0 0 0 0]{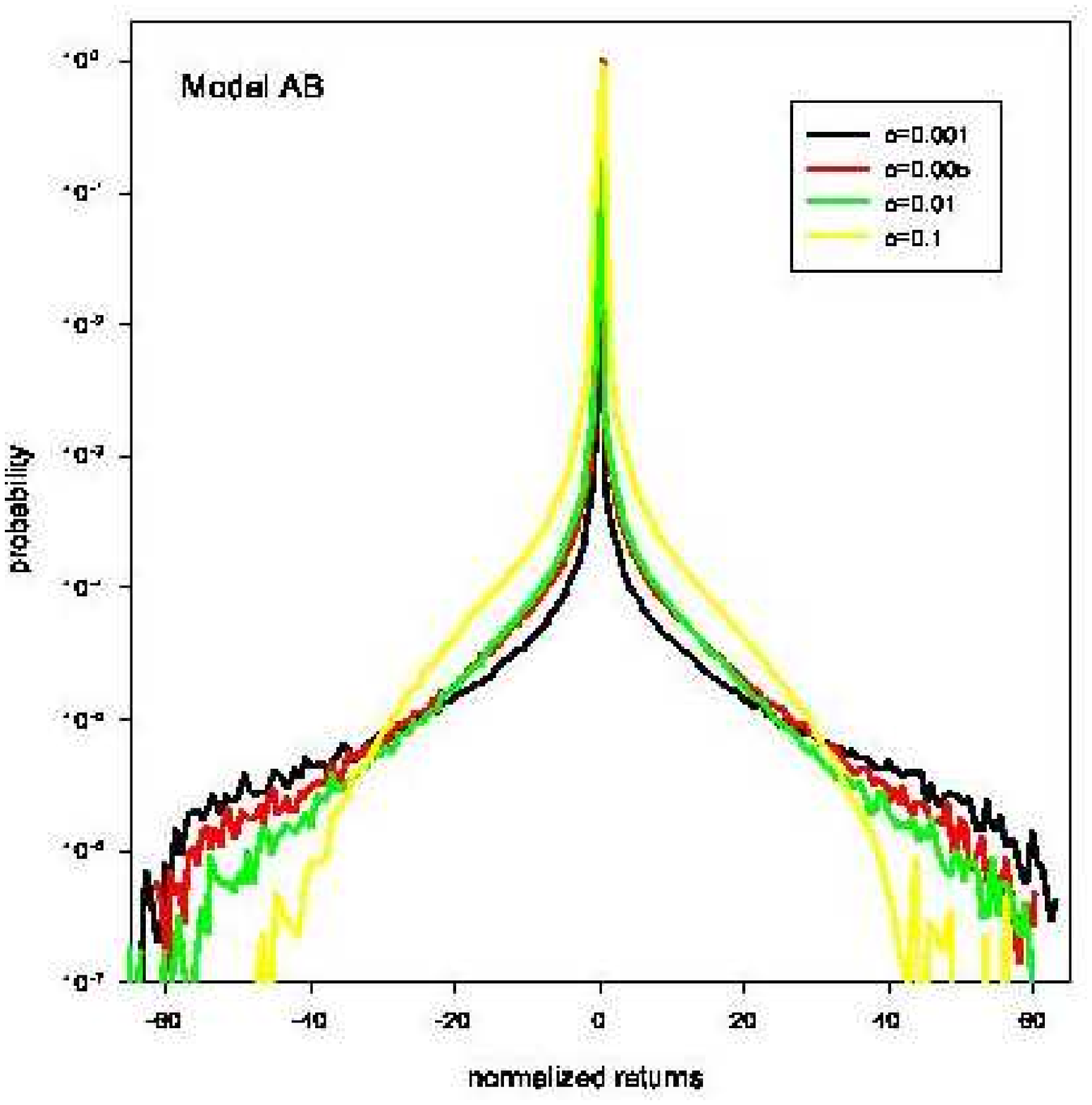} 
    \end{center} 
    \caption{Distribution of normalized returns for 
both models considered. Left panel: model A. Right panel: model AB. Notice
really huge returns observed if $b$ is small enough and two symmetric dips
around the zero-return peak in case of the curve associated to the model
A. Tails of the distributions with $b>0.1$  decay accordingly to the power-law with exponent  about 1.8} 
    \label{fig:distribution} 
\end{figure} 
 
On Fig. \ref{fig:distribution} we show distributions of normalized returns.
The normalization means dividing each return by the standard deviation. 
The synchronized decisions of clusters result in huge returns and what follows
in fat  wings of the return distribution. However the distribution arisen
from the model A points at the absence of small returns also. Close to the zero-return peak, symmetrically on its both sides, we observe two minimums. But with $b \ge 0.1$ both models lead to distributions of returns hardly distinguishable from the CB model distribution. It seems that effect on a price caused by actions of more than 40 independent groups of investors is similar to that ones which are undertaken by the whole 400 independent investors possible to exists on $200\times 200$ lattice.

\section{ Polish stock market  and imitation}  
 
\begin{figure} 
    \begin{center} 
        \includegraphics[width=0.49\textwidth, trim= 0 0 0 0]{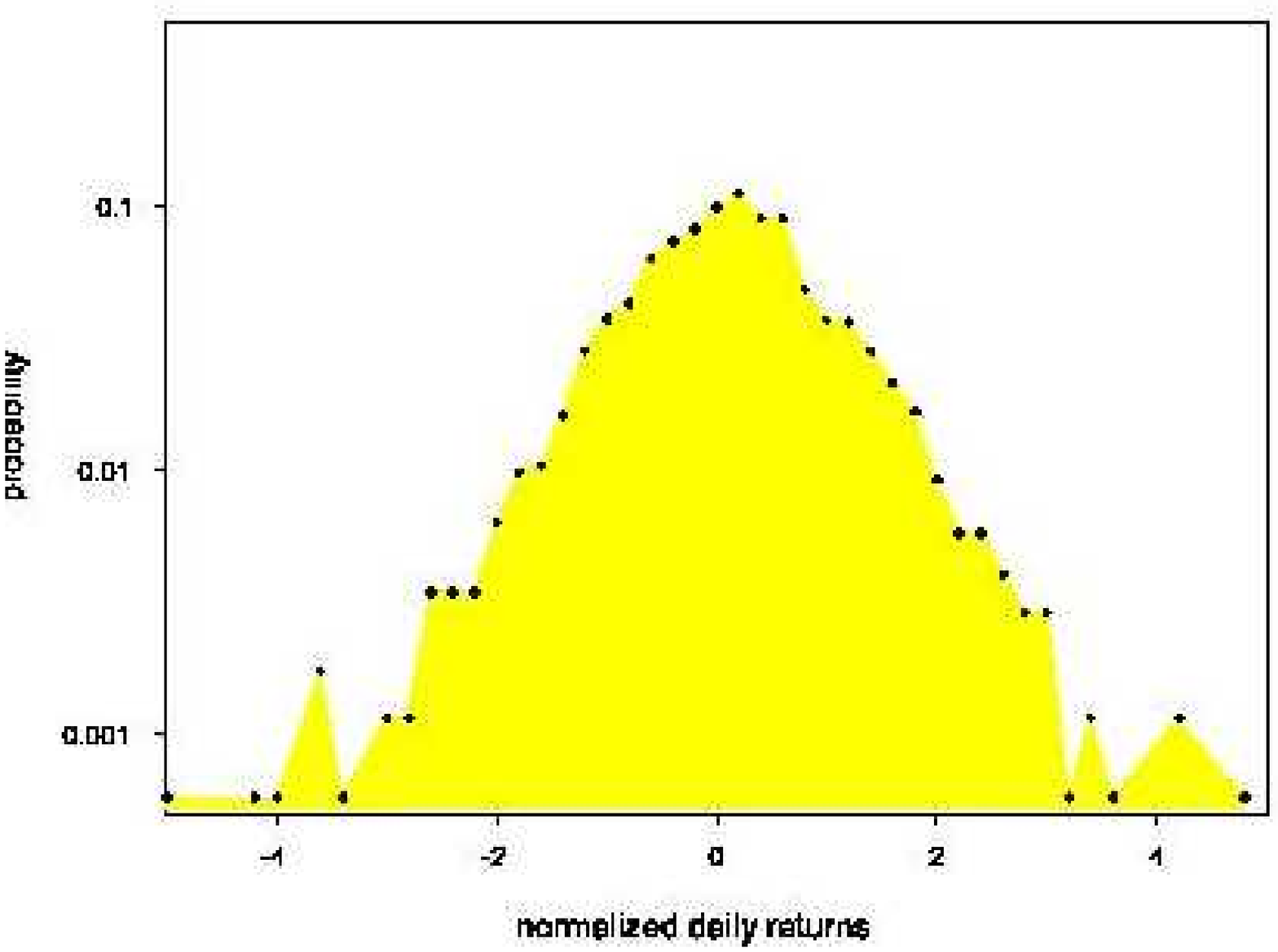} 
        \includegraphics[width=0.49\textwidth, trim= 0 0 0 0]{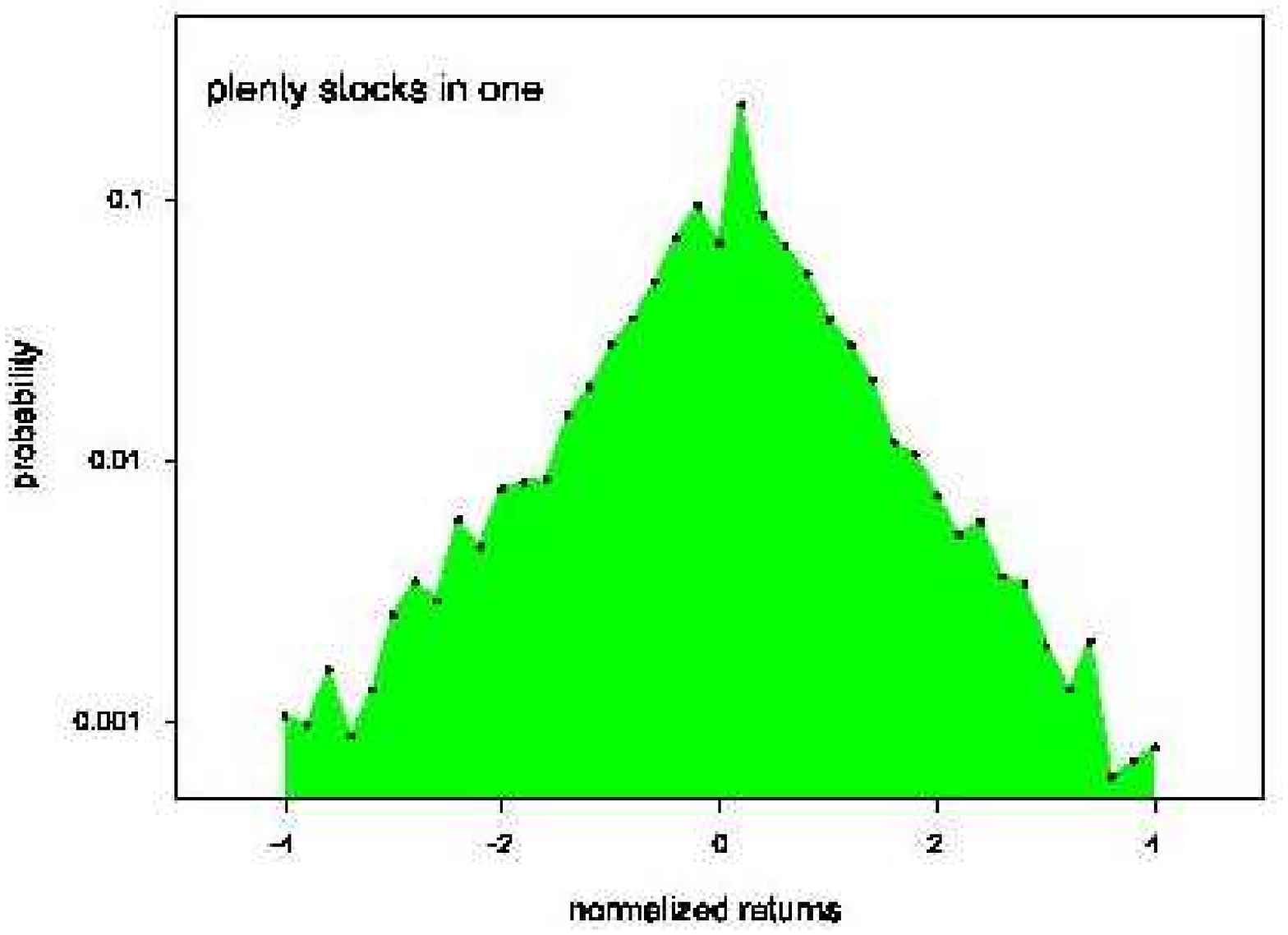} 
        \end{center} 
\caption{Left panel: distribution of returns of Warsaw stock index, right
panel: distribution of returns of stocks classified as irregular} 
\label{fig:index} 
\end{figure}

\begin{figure} 
    \begin{center} 
        \includegraphics[width=0.49\textwidth, trim= 0 0 0 0]{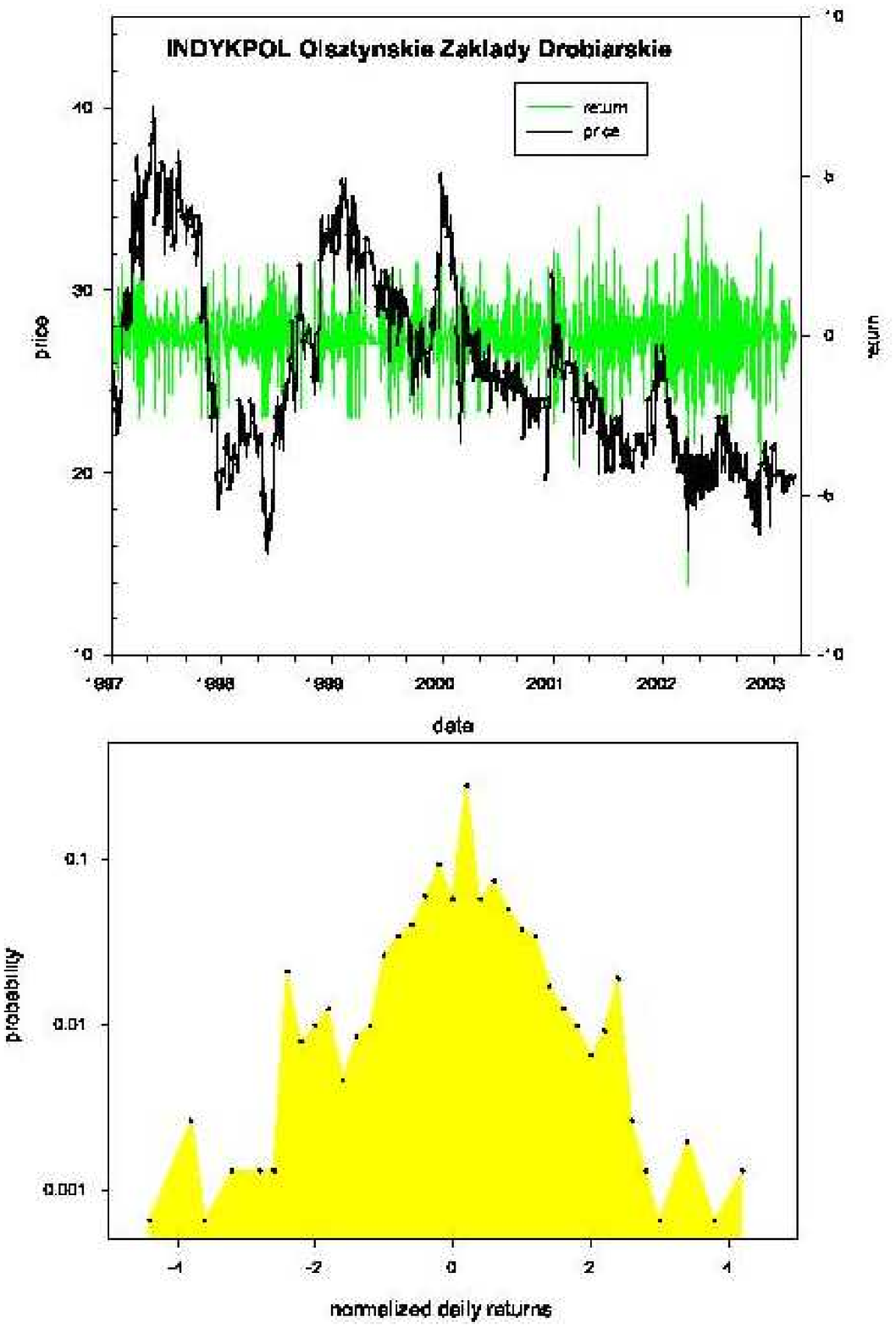} 
        \includegraphics[width=0.43\textwidth, trim= 0 0 0 0]{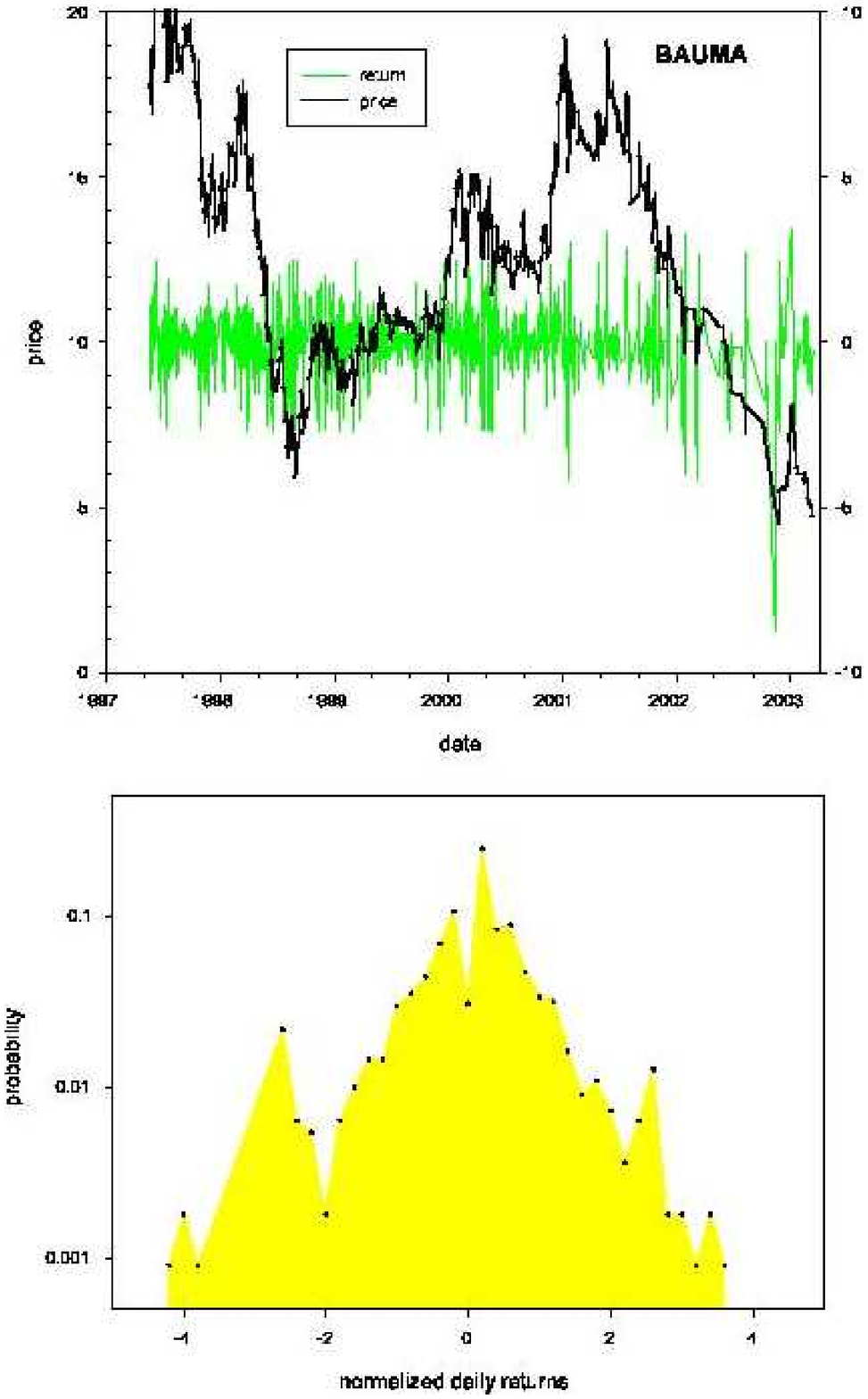} 
        \end{center} 
\caption{Time series of stocks which lead to  distribution of returns irregular. Both firms are small. DFA gives: 0.329 for Indykpol and 0.408 for Bauma } 
\label{fig:stocks} 
\end{figure}

We have examined 73 series of prices of stocks from Polish stock market to
find signs of the model A: dips around the zero-return peak, or the model
AB: fat tails. We analysed the date which were collected during last six  or seven years. The Warsaw Stock Exchange regulations on the minimal price
change imply that the size of a histogram bin has to be carefully selected.
In the following figures we plot distributions of logarithmic returns, namely,
return(t)= log price(t) -log price (t-1), which are normalized by their standard
deviation. The minimal price change rule leads to the separation from the return zero less than 0.064. Therefore we decided that  the value 0.20 for the size for a histogram bin is proper. Accordingly to the distribution shape the stocks have been divided into two groups: regular - with the  distribution similar to the distribution
of WIG- the Warsaw Stock Exchange market index, and others, see Fig.~\ref{fig:index}. Let us point at the minimum on the left side of the zero-return peak. To identify stocks with irregular distribution we searched for the zero-returns properties and fat tails. Series classified as possible examples which exhibit properties of the model A and the model AB are collected in Fig.~\ref{fig:stocks}. Together we show time series of these stocks. These stocks represent weak firms: low capital, low liquidity, often at the edge of the bankruptcy. The DFA method \cite{Peng} for quantification of long-termed correlation in time-series provides the crossover to the antipersistency behavior for long time distances. 40 stocks of all 73 considered by us are found with the corresponding exponent $\alpha <0.48$. Among them 19 shows  $\alpha <0.40$.

\section{Closing} 
Presented results are only preliminary and further investigations should be undertaken.
The proposed microdynamics  leads to the non-realistic global
effect in a sense that each price change is accepted. In real markets there
are rules preventing from unlimited large or infinitely  small changes. 
Nevertheless, the observed on the Polish  market time series  justify the modifications introduced by us to  the original CB-model.  Specially when emerging markets are considered then the assumption about imitating others in place of  undertaking own reasoning seems to be legalized. However one must be careful in a quantitative conclusion suggested when comparing the model A distribution to a real series distribution that a stock is driven by not more than a few trees of investors. 

There are some natural possibilities for further model development. The existing extension of the CB model such as adjusting activity with respect to the latest price movements \cite{StaufferOliveiraBernardes} offers the straightforward way. The next possibility is to inject imitating agents, agents of the class B, into different from the CB model dynamics. For example,  it should be interesting to consider the Lux-Marchesi model \cite{LuxMarchesi} with speculative investors divided into imitation trees. Finally, one should consider switching imitating agents into overconfident investors, which are know to be present on well developed markets \cite{overconfidence}. Such investors being overconfident about own reasoning, understand better a market than other investors. Therefore each of overconfident investors can predict with certainty what some other investor will do at the next time step. Having such a knowledge he modifies his present decision. Thus, he buys now if the other investor will buy, he sells if the other investor will sell. Simulations with described microdynamics are performed and soon will  be presented.

{\bf Acknowledgement\\} 
We are grateful to Professor Ditrich Stauffer for the source code of Cont-Bouchaud
model.  
This work is supported by  Gda\'nsk University: BW 5400-5-0014-3. 
The simulations, partially, have been performed in TASK- Academic Computer
Center in Gdansk.

\end{document}